\documentclass[a4paper,11pt]{article}
\usepackage{pos}
\usepackage{graphicx}
\usepackage{slashed}
\title{Electromagnetic  neutrino:
The theory and reactor and solar bounds on neutrino magnetic moments}
\ShortTitle{Electromagnetic neutrinos: The theory and bounds from scattering experiments}

\author*[a]{Alexander Studenikin}

\affiliation[a]{Department of Theoretical Physics, Moscow State University,\\
  119992 Moscow, Russia}

\affiliation[b]{Dzhelepov Laboratory of Nuclear Problems, Joint Institute for Nuclear Research,\\
141980 Dubna, Russia}

\emailAdd{studenik@srd.sinp.msu.ru}
\emailAdd{a-studenik@yandex.ru}

\abstract{A brief overview of the electromagnetic properties of neutrinos is presented with a discussion of the most important fundamental aspects of the problem. Then using data from the ground-based reactor and solar neutrino-electron scattering experiments the best upper bounds on neutrino effective magnetic moments are discussed.}

\FullConference{%
  *** The European Physical Society Conference on High Energy Physics (EPS-HEP2021), ***\\
  *** 26-30 July 2021 ***\\
  *** Online conference, jointly organized by Universität Hamburg and the research center DESY ***
}

\begin{document}
\maketitle

\section{The theory}
It is usually assumed that the entirety of the electromagnetic properties of neutrinos are embodied by the structure of the amplitude corresponding to the Feynman diagramme shown in Fig. 1 that describes the interaction of a neutrino with a real photon.
\begin{figure}[h]
\begin{center}
\includegraphics[width=10pc]{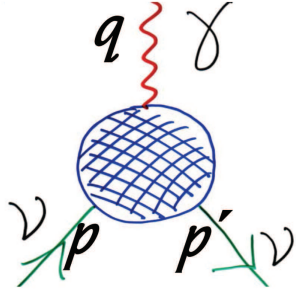}
\caption{\label{label}A neutrino effective one-photon coupling.}
\end{center}
\end{figure}
Two incoming and outgoing lines represent initial and final neutrino states and the third line stands for the real photon. These three lines are connected by an effective vertex, which in general contains the whole set of the neutrino electromagnetic characteristics. The diagramme in Fig. 1 corresponds to the one-photon approximation for the electromagnetic interactions of a neutrino field $\nu (x) $ that can be described by the effective interaction Hamiltonian
\begin{equation}
H_{em} (x)= j_{\mu}(x) A^{\nu} (x) = {\bar \nu(x)} \Lambda _{\mu} \nu (x) A^{\mu} (x),
\end{equation}
where $j_{\mu}(x)$ is the neutrino effective electromagnetic current and $\Lambda_{\mu}$ is a matrix in the spinor space. Considering the neutrinos as free particles with the Fourier expansion of the free Dirac fields, for the amplitude corresponding to the diagramme in Fig. 1 it is possible to get (see \cite{Giunti:2014ixa} and \cite{Nowakowski:2004cv} for the detailed derivations)

\begin{equation}
\langle \nu(p_f)|j_\mu (0)|\nu(p_i)\rangle=\bar{u} (p_f)\Lambda_{\mu}(q)u(p_i),
\end{equation}
where $q=p_i-p_f$.

In the most general form the neutrino electromagnetic vertex function $\Lambda_{\mu}^{ij}(q)$ can be expressed \cite{Giunti:2014ixa} in terms of four form factors
\begin{equation}\label{Lambda}
\Lambda_{\mu}^{ij}(q) =  \left( \gamma_{\mu} - q_{\mu}
\slashed{q}/q^{2} \right) \left[ f_{Q}^{ij}(q^{2}) + f_{A}^{ij}(q^{2})
q^{2} \gamma_{5} \right]
 - i \sigma_{\mu\nu} q^{\nu} \left[ f_{M}^{ij}(q^{2}) +
i f_{E}^{ij}(q^{2}) \gamma_{5} \right] ,
\end{equation}
where $\Lambda_{\mu}(q)$ and form factors $f_{Q,A,M,E}(q^2)$ are $3\times 3$ matrices in  the space of massive neutrinos.
Note that in the derivation of the decomposition (\ref{Lambda}) the demands followed from the Lorentz-invariance and electromagnetic gauge invariance are taken into account.

In the case of coupling with a real photon ($q^2=0$) the form factors $f(q^2)$ provide  four sets of neutrino electromagnetic characteristics: 1) the electric millicharges $q_{ij}=f_{Q}^{ij}(0)$,
2) the dipole magnetic moments $\mu_{ij}=f_{M}^{ij}(0)$, 3) the dipole electric moments $\epsilon_{ij}=f_{E}^{ij}(0)$ and
4) the anapole moments $a_{ij}=f_{A}^{ij}(0)$.

The expression (\ref{Lambda})  for $\Lambda_{\mu}^{ij}(q)$ is applicable for Dirac and Majorana neutrinos. However, a Majorana neutrino does not have diagonal electric charge and dipole magnetic and electric form factors, only a diagonal anapole form factor can be nonzero. At the same time, a Majorana neutrino can also have nonzero off-diagonal (transition) form factors.

If one considers the case of equal masses for the initial and final neutrinos, $m_i=m_f$, then the decomposition of the neutrino electromagnetic  vertex function reduces to
\begin{equation}\label{Lambda_mm}
\Lambda_{\mu}^{ii}(q) =  f_{Q}^{ii}(q^2) \gamma_{\mu}  \nonumber
 - i \sigma_{\mu\nu} q^{\nu} \left[ f_{M}^{ii}(q^{2}) +
i f_{E}^{ii}(q^{2}) \gamma_{5} \right] + f_{A}^{ii}(q^{2})\left( \gamma_{\mu} - q_{\mu}
\slashed{q}/q^{2} \right)\gamma_{5}.
\end{equation}

It is also interesting to consider neutrino electromagnetic properties for the case of massless neutrinos  of the Standard Model when neutrinos are described by two-component left-handed Weyl spinors. In this case neutrinos have only one form factor which is equal to the difference of the charge $f_Q (q^2)$ and anapole $f_A(q^2)$ form factors, and the electromagnetic vertex function is given by
\begin{equation}\label{Lambda_mm}
\Lambda_{\mu}(q) =(\gamma_{\mu}- q_{\mu}
\slashed{q}/q^{2}) f(q^2),
\ \ \ \ \ f(q^2)= f_Q (q^2)- f_A(q^2).
\end{equation}
From these expressions one can expect  that at least from the phenomenological point of view it is not possible to treat consequences of neutrino nonzero electric charge $f_Q (q^2)$ and anapole $f_A (q^2)$ form factors separately. This case can approximate the relation between these two form factors that arises for an ultrarelativistic massive neutrinos.

Taking into account the above-mentioned feature of the anapole form factor $f_A (q^2)$ (the ambiguity of its allocation against the charge form factor $f_Q (q^2)$  which manifests itself even in the Standard Model), an alternative decomposition of the electromagnetic vertex function $\Lambda_{\mu}(q)$ has been proposed in \cite{Dubovik:1996gx}. In \cite{Dubovik:1996gx} the toroidal dipole form factor $f_{T}(q^2)$ is introduced as a characteristic of the neutrino instead of the anapole form factor $f_A (q^2)$. In this case, the neutrino electromagnetic vertex can be written in the so-called toroidal parametrization
\begin{equation}\label{Lambda_toroidal}
\Lambda_{\mu}^{ij}(q) =  f_{Q}^{ij}(q^2)\gamma_{\mu}  \nonumber
 - i \sigma_{\mu\nu} q^{\nu} \left[ f_{M}^{ij}(q^{2}) +
i f_{E}^{ij}(q^{2}) \gamma_{5} \right] + if_T ^{ij} (q^2) \epsilon_{\mu \nu \lambda \rho} P^\nu q^\lambda \gamma ^\rho,
\end{equation}
where $P=p_i + p_f$ and $\epsilon_{\mu \nu \lambda \rho}$ is the Levi-Civitta unit antisymmetric tensor. The toroidal parametrisation of the neutrino vertex has a more clear physical interpretation than the anapole one, because it provides a one-to-one correspondence between the form factors and the multipole moments in expansion of electromagnetic fields. The corresponding toroidal dipole moment $f_T (q^2=0)$, for the first time introduced in \cite{Zeldovich}, is more convenient than the anapole moment $f_A (q^2=0)$ for the description of T-invariant interactions with nonconcervation of the P and C symmetries.

From the identity
\begin{equation}
{\bar u}(p_f)\Big[(m_i-m_f)\sigma _{\mu\nu}g^\mu +(g^2\gamma_\mu -{\slashed q}q_\mu)-\epsilon_{\mu\nu\lambda\rho} P^{\nu}q^{\lambda}\gamma^{\rho} \gamma_5\Big]\gamma_5 u_i (p_i)=0
\end{equation}
it follows \cite{Dubovik:1996gx} that in the static limit the toroidal $f_T (q^2=0)$ and anapole  $f_A (q^2=0)$ dipole moments coincide when the masses of the initial and final neutrino states are equal to each other, $m_i=m_f$.

\section{Neutrino electromagnetic properties in scattering experiments}
The possible electromagnetic characteristics of neutrinos can manifest themselves in astrophysical environments, where neutrinos propagate in strong magnetic fields and dense matter, and also in ground-based laboratory measurements of neutrinos fluxes from various sources. So far, there are no any evidenced in favour of neutrino nonzero electromagnetic properties either from laboratory measurements of neutrinos from ground-based sources or from observations of neutrinos from astrophysical sources. Only constraints (the upper bounds on neutrino electromagnetic characteristics) are obtained in different experiments. The available constraints are discussed in the review paper \cite{Giunti:2014ixa} (see also \cite{Studenikin:2013my}, \cite{Parada:2019gvy}, \cite{Studenikin:2020nky}, \cite{Brdar:2020quo} and \cite{Studenikin_TAUP_2021} for the latest developments and progress in this field).

A widely used method to probe the neutrino electromagnetic properties is based on  the direct measurements of low-energy elastic (anti)neutrino-electron scattering in reactor, accelerator, and solar neutrino experiments. Possible nonzero electromagnetic characteristics of neutrinos, such as the electric millicharges, charge radii, dipole magnetic and electric moments, and anapole (toroidal) moments can provide additional contributions to the scattering of neutrinos on a target which are measured in the corresponding experiments.

A general strategy of such experiments consists in determining deviations of the scattering cross section differential with respect to the energy transfer from the value predicted by the Standard Model of the electroweak interaction. The desired goal of such experiments may be, firstly, to obtain information about the magnitude of the contributions of electromagnetic interactions of neutrinos to the cross sections. Then this information (a modification of the scattering cross-section confirmed in a specific experiment due to electromagnetic interactions of neutrinos) can be used to study the patterns of neutrino propagation in various media or to analyze data from various other experiments. Within the framework of this approach, we are not interested in the nature of the occurrence of nontrivial electromagnetic properties of neutrinos. In this case, only restrictions on the numerical values of specific electromagnetic characteristics of neutrinos extracted from the data of the neutrino scattering experiment are used. The question of the origin of the nonzero electromagnetic properties of neutrinos, that is, about a more fundamentally theoretical model predicting the nonzero properties of neutrinos, remains behind the scenes.

The second main purpose of neutrino scattering experiments may be to compare the obtained experimental data on the cross section (more precisely, from the obtained upper bounds of the possible electromagnetic contribution to the cross section) with the predictions of a more general theoretical model that provides nonzero electromagnetic characteristics of neutrinos from the first fundamental principles. In this context, we can say that the study of the electromagnetic properties of neutrinos opens a window into new physics.

The most consistent approach to the theoretical description of the electromagnetic properties of neutrinos involves the initial introduction of nonzero electromagnetic characteristics for the mass states of neutrinos. At the same time, since in neutrino scattering experiments flavour neutrinos are registered in detectors, it is necessary to translate electromagnetic properties from the mass into the flavour neutrino basis. Therefore, due to the neutrino mixing and oscillations along the neutrino path from the source to the detector
the observed (constrained) neutrino electromagnetic characteristics depends on the neutrino flavour composition in the detector. The recent and most comprehensive study of  neutrino electromagnetic properties in the neutrino electron scattering with account for neutrino mixing and oscillations can be found in \cite{Kouzakov:2017hbc}.

Consider the most stringent constraints on the effective neutrino
magnetic moments that are obtained with the reactor antineutrinos:
$\mu_{\nu} < 2.9 \times 10^{-11} \mu_{B}$ (GEMMA Collaboration \cite{GEMMA:2012}),
and solar neutrinos:
${\mu}_{\nu}\leq 2.8 \times
10^{-11} \mu _B$ (Borexino Collaboration \cite{Borexino:2017fbd}).
Both these constraints are obtained with investigations of the elastic scattering of a flavour neutrino $\nu_l$ (or an antineutrino $ { \bar \nu}_l$ ) on an electron at rest:
$\nu_{l} + e^{-} \to \nu_{l} + e^{-}, \ \ l=e, \mu, \tau $.
 There are two contributions, one from the Standard Model weak interaction and another one from the neutrino magnetic moment interaction,  to the electron neutrino cross section  \cite{Vogel:1989iv},
\begin{equation}\label{dsigma}
\frac{d\sigma_{\nu_{l}e^{-}}}{dT_{e}}
=
\left(\frac{d\sigma_{\nu_{l}e^{-}}}{dT_{e}}\right)_{SM}
+
\left(\frac{d\sigma_{\nu_{l}e^{-}}}{dT_{e}}\right)_{\mu}.
\end{equation}
  The weak-interaction cross section is
\begin{equation}
\left(\frac{d\sigma_{\nu_{l}e^{-}}}{dT_{e}}\right)_{SM}
= \frac{G^{2}_F m_{e}}{2\pi}
\bigg\{ (g_{V}^{\nu_{l}} + g_{A}^{\nu_{l}})^{2}
+
(g_{V}^{\nu_{l}} - g_{A}^{\nu_{l}})^{2}
\left(1-\frac{T_{e}}{E_{\nu}}\right)^{2}
+
\left[ (g_{A}^{\nu_{l}})^{2} - (g_{V}^{\nu_{l}})^{2} \right]
\frac{m_{e}T_{e}}{E_{\nu}^{2}}
\bigg\},
\label{dsigmaSM}
\end{equation}
with the standard coupling constants $g_{V}$ and $g_{A}$ given by:
$g_{V}^{\nu_{e}}
=
2\sin^{2} \theta_{W} + 1/2
, g_{A}^{\nu_{e}}
=
1/2,
\ \ \ g_{V}^{\nu_{\mu,\tau}}
=
2\sin^{2} \theta_{W} - 1/2,
\ \ \ g_{A}^{\nu_{\mu,\tau}}
=
- 1/2$.
The cross section depends on the initial neutrino energy $E_\nu$ and also contains the electron recoil energy $T_e$. For antineutrinos one must substitute
$g_A \to -g_A$.
If to account that the born in the source flavour neutrino $|\nu_l\rangle$ arrives to the detector in the flavour state given by
\begin{equation}
|\nu_l(L)\rangle=\sum_{k=1}^{3}U^{\ast}_{lk}e^ {-\frac{m^2_k}{2E_\nu}L}|\nu_k\rangle,
\end{equation}
then the neutrino magnetic-moment contribution to the cross section is given by
\begin{equation}
\left(\frac{d\sigma_{\nu_{l}e^{-}}}{dT_{e}}\right)_{\mu}
=
\frac{\pi\alpha^{2}}{m_{e}^{2}}
\left(\frac{1}{T_{e}}-\frac{1}{E_{\nu}}\right)
\left(\frac{\mu_{\nu_{l}}}{\mu_B}\right)^{2}
,
\label{dsigmamu}
\end{equation}
here $\mu_B$ is the Bohr magneton. The cross section contains the effective magnetic moment $\mu_{\nu_{l}}$ \cite{Giunti:2014ixa}, \cite{Kouzakov:2017hbc}
\begin{equation}
{\mu_{\nu_{l}}}^{2}(L,E_{\nu})
=
\sum_{j}
\left|
\sum_{k}
U_{l k}^{*}
e^{- i \Delta{m}^{2}_{kj} L / 2 E_{\nu} }
\left(
\mu_{jk} - i \epsilon_{jk}
\right)
\right|^{2}
\label{mueff}
\end{equation}
that indeed receives equal contributions from the neutrino electric and magnetic dipole moments, both the diagonal and transition, which are given by the static values of the corresponding form factors,
$\mu_{jk} = f_M^{jk}(q^2=0), \ \ \ \ \epsilon_{jk} = f_E^{jk}(q^2=0)$.

It is just straightforward that scattering experiments based on detection of neutrinos arriving from different distances probe (or constraint) different combinations of the fundamental magnetic moments. Now for simplicity we omit possible contributions from the dipole electric moments. In the reactor short-baseline experiments (for instance, in the GEMMA experiment) one studies (and constrains) the effective magnetic moments in the flavor basis. Whereas in the case of long-baseline experiments (such as the Borexino experiment with the solar neutrinos) a more convenient interpreting the results is based on effective  magnetic moments in the fundamental mass basis, rather than in the flavour basis. These should be accounted for when one compares quite similar values for effective magnetic moments obtained form the GEMMA and Borexino experiments.

Note that with the inclusion \cite{Studenikin:2013my} of the effect from possible nonzero neutrino millicharge to the analysis of the GEMMA collaboration data on the antineutrino-electron scattering provides the most severe reactor upper bound on the neutrino millicharge: $q_{\nu}< 1.5 \times 10^{-12} e_0 $. For a detailed and recent discussion on the main processes of the electromagnetic interaction of neutrinos in astrophysics and the corresponding limitations on millicharges and effective magnetic moments of neutrinos see \cite{Studenikin_TAUP_2021}.

The author is thankful to Konstantin Kouzakov for useful discussions. The work is supported by the Interdisciplinary Scientific and Educational School of Moscow University “Fundamental and Applied Space Research” and by the Russian Foundation for Basic Research under grant No. 20-52-53022-GFEN-a.

\end{document}